\documentclass[conference]{IEEEtran}
\usepackage{tikz}	 
\usepackage{adjustbox} 
\usetikzlibrary{arrows.meta} 
\usetikzlibrary{positioning,calc,shapes}
\usetikzlibrary{matrix,calc} 

\usepackage{enumitem}	

\usepackage[ansinew]{inputenc}
\usepackage{graphicx}
\usepackage{epsfig}
\usepackage{amsmath}    
\usepackage{amssymb}
\usepackage{subcaption}
\usepackage[all]{xy}
\usepackage{multirow}
\usepackage{tikz}
\usetikzlibrary{snakes,matrix,trees,arrows}

\usepackage[a4paper,left=1.8cm,right=1.8cm,top=2.4cm,bottom=2.4cm]{geometry}

\usepackage{float} 
\usepackage{color}
\usepackage{xcolor}
\usepackage{listings}

\makeatletter

\def\zapcolorreset{\let\reset@color\relax\ignorespaces}
\def\colorrows#1{\noalign{\aftergroup\zapcolorreset#1}\ignorespaces}

\makeatother

\usepackage{acronym}
\acrodef{lln}[LLNs]{Low-Power and Lossy Networks}
\acrodef{rpl}[RPL]{IPv6 Routing Protocol for Low-Power and Lossy Networks}
\acrodef{mop}[MOP]{Mode of Operation}

\acrodef{mp2p}[MP2P]{multipoint-to-point}
\acrodef{p2p}[P2P]{point-to-point}
\acrodef{p2mp}[P2MP]{point-to-multipoint}

\acrodef{dodag}[DODAG]{Destination-Oriented DAG}
\acrodef{dio}[DIO]{DODAG Information Object}
\acrodef{dao}[DAO]{Destination Advertisement Object}
\acrodef{dis}[DIS]{DODAG Information Solicitation}

\acrodef{ai}[AI]{Artificial Intelligence}
\acrodef{iot}[IoT]{Internet-of-Things}
\acrodef{sdn}[SDN]{Software-Defined Networking}

\begin{document}
\renewcommand\texteuro{FIXME} 

\title{Implementation of RPL in OMNeT++}


\author{\IEEEauthorblockN{Hedayat~Hosseini}
\IEEEauthorblockA{Computer Engineering and\\ 
Information Technology Department\\ 
Amirkabir University of Technology\\ 
(Tehran Polytechnic)\\ 
Tehran, Iran\\ 
e-mail: h.hosseini@aut.ac.ir}
\and
\IEEEauthorblockN{Elisa Rojas}
\IEEEauthorblockA{Universidad de Alcal\'a\\ 
Departamento de Automatica\\ 
28805, Alcala de Henares, Spain\\ 
e-mail: elisa.rojas@uah.es}
\and
\IEEEauthorblockN{David Carrascal}
\IEEEauthorblockA{Universidad de Alcal\'a\\ 
Departamento de Automatica\\ 
28805, Alcala de Henares, Spain\\ 
e-mail: david.carrascal@uah.es}
}

\markboth{Version 1.00}%
{Hosseini \MakeLowercase{\textit{et al.}}: Implementation of RPL in OMNeT++}

\maketitle
\thispagestyle{plain}
\pagestyle{plain}


\begin{abstract}
The growth and evolution of Internet of Things (IoT) is now of paramount importance for next-generation networks, including the upcoming 6G. In particular, there is a set of constrained IoT nodes that comprise the Low-Power and Lossy Networks (LLNs), which have very particular requirements. The current standard for routing in those networks is RPL, which was defined less than a decade ago and still needs improvements in terms of scalability or integration with other networks. Many researchers currently need an implementation of RPL to evaluate their works and, for that reason, among others, we implemented it in the OMNeT++ simulator. The results of this implementation show that is an easy way to check prototypes in their very initial develop phases, and its code is publicly available for the research community. 
\end{abstract}

\begin{IEEEkeywords}
Network simulation, OMNeT++, IoT, RPL, 6G 
\end{IEEEkeywords}

\IEEEpeerreviewmaketitle

\section{Introduction}
\label{introduction}

The future sixth generation of mobile/wireless networks, also known as 6G, is envisioned to cause a tremendous growth in the number of connected devices (that is, \ac{iot} nodes) in order to achieve diverse applications of smart environments, fostered by \ac{ai}, such as Agriculture 4.0 or Industry 4.0~\cite{Tomkos20}. 
However, still many challenges need to be tackled, like the seamless integration of \ac{iot} in virtualized and programmable environments required by 6G~\cite{Carrascal20}.

\ac{rpl}~\cite{rfc6550} is an standard for routing in~\ac{lln}. \acp{lln} consist of a set of constrained \ac{iot} nodes interconnected by limited links. The constraint is due to processing, memory, and sometimes energy resources of the nodes and high loss rates, low data rates, and instability of the links~\cite{rfc6550}. 
Currently, there are many efforts to align RPL to the requirements of these 6G networks and, therefore, there are many proposals enhancing RPL in different ways. 

In our specific case, our research team wanted to implement an alternative protocol for \acp{lln}. To evaluate it and compare it with \ac{rpl}, we decided to first implement \ac{rpl} in OMNeT++. This served as an initial step to prove our designed prototype, and we hope it also helps future researchers trying to design additional protocols and enhanced versions of \ac{rpl}.

Our article is structured as follows. 
In Section~\ref{rpl}, we first provide a summarized overview of \ac{rpl}, explaining its main \acp{mop} and its design implications. 
Afterwards, in Section~\ref{platforms} we analyze the diverse network simulators and emulators, and explain our reason to choose OMNeT++ as the implementation platform for our initial prototype. 
In Sections~\ref{implementation} and~\ref{installation} we describe the main characteristics of the implementation, and how to install it, respectively. 
In Section~\ref{results} we examine the obtained results and expected impact of our implementation. 
Finally, we conclude the article Section~\ref{conclusion}, providing some research directions as well.

\section{RPL overview}
\label{rpl}

In order to explain the design decisions made along the implementation of \ac{rpl} in the OMNeT++ simulator, we will first provide a quick overview about the protocol.

Each~\ac{rpl}-based network can operate in one of three modes illustrated in Table~\ref{table:mopEncoding}~\cite{rfc6550}. Upward routes are established based on a common routing used in the all \ac{mop}s in the standard. 
These routes are used for \ac{mp2p} traffic type. 
Therefore, we could say that \ac{mop}s are defined in terms of whether the Downward routes are used to enable a network to route \ac{p2mp} and \ac{p2p} traffic types or not, and also based on how \ac{rpl} populates the routing/source routing tables of each node in the network.

\begin{table}[!h]
	\centering
	\caption{\ac{mop} encoding description in~\ac{rpl}~\cite{rfc6550}}
	\label{table:mopEncoding}
	\begin{tabular}{|l|l|}
		\hline
		\textbf{\ac{mop} \#}& \textbf{Description} of \ac{mop}\\
		\hline
		0 & No Downward routes maintained by RPL\\
		1 & Non-Storing Mode of Operation\\
		2 & Storing Mode of Operation with no multicast support\\
		3 & Storing Mode of Operation with multicast support\\
		Other values & Unassigned\\
		\hline
	\end{tabular}
\end{table}

\subsection{MOP 0}
When \ac{rpl} uses \ac{mop} 0, all nodes can only use Upward routes and \ac{mp2p} traffic type from a node to the \ac{dodag} root along the \ac{dodag}. Therefore, all non-root nodes can send their traffic, called multi-point, to a root node, called point, in the network. Since in this \ac{mop} routing is unidirectional,  \ac{p2p} traffic type cannot be used to communicate two non-root nodes with each other (particularly when the nodes are not neighbors of each other). In this regard, there are some schemes to improve and optimize \ac{p2p} traffic type in \ac{rpl}~\cite{rfc6550}. 

\begin{figure}[hbt]
    \centering
    \begin{adjustbox}{width=0.48\textwidth}
        \tikzstyle{myNetworkNode}=[circle,draw=blue!50,fill=blue!20,thick,inner sep=0pt,minimum size=20mm]
        \tikzstyle{myEdge}=[]
        \tikzstyle{myCalloutUp}=[ellipse callout, fill=blue!20,opacity=.8, yshift=1.5cm]
        \tikzstyle{myCalloutDown}=[ellipse callout, fill=blue!20,opacity=.8, yshift=-1.5cm]
        \tikzstyle{myAddressTable}=[]
        \begin{tikzpicture}
            \node[myNetworkNode] (1) at (0,0) {\scalebox{4}{1}};
            \node[myNetworkNode] (2) at (-8,-8) {\scalebox{4}{2}};
            \node[myNetworkNode] (3) at (8,-8) {\scalebox{4}{3}};
            \node[myNetworkNode] (4) at (-12,-16) {\scalebox{4}{4}};
            \node[myNetworkNode] (5) at (-4,-16) {\scalebox{4}{5}};
            \node[myNetworkNode] (6) at (12,-16) {\scalebox{4}{6}};
            \node[myNetworkNode] (7) at (16,-24) {\scalebox{4}{7}};

            \node[myAddressTable] at (-8,-30) {
                \scalebox{4}{                
                    \begin{tabular}{|c|c|c|}
                        \hline
                        \textbf{Node ID} & \textbf{Link Local Address} & \textbf{Global Address} \\ 
                        \hline
                        1 & fe80::8aa:ff:fe00:1 & fd00::8aa:ff:fe00:1 \\ 
                        \hline
                        2 & fe80::8aa:ff:fe00:2 & fd00::8aa:ff:fe00:2 \\ 
                        \hline
                        3 & fe80::8aa:ff:fe00:3 & fd00::8aa:ff:fe00:3 \\ 
                        \hline
                        4 & fe80::8aa:ff:fe00:4 & fd00::8aa:ff:fe00:4 \\ 
                        \hline
                        5 & fe80::8aa:ff:fe00:5 & fd00::8aa:ff:fe00:5 \\ 
                        \hline
                        6 & fe80::8aa:ff:fe00:6 & fd00::8aa:ff:fe00:6 \\ 
                        \hline
                        7 & fe80::8aa:ff:fe00:7 & fd00::8aa:ff:fe00:7 \\ 
                        \hline
                    \end{tabular}
                }
            };

            \node at (0,3){
                \scalebox{3.5}{
                    \begin{tabular}{|c|c|}
                        \hline
                        \textbf{Destination} & \textbf{Next Hop} \\ 
                        \hline
                    \end{tabular}
                }
            };
            \node at (-12,-6){
                \scalebox{3.5}{
                    \begin{tabular}{|c|c|}
                        \hline
                        \textbf{Destination} & \textbf{Next Hop} \\ 
                        \hline
                        Default Route & fe80::8aa:ff:fe00:1\\  
                        \hline
                    \end{tabular}
                }
            };
            \node at (12,-6){
                \scalebox{3.5}{
                    \begin{tabular}{|c|c|}
                        \hline
                        \textbf{Destination} & \textbf{Next Hop} \\ 
                        \hline
                        Default Route & fe80::8aa:ff:fe00:1\\  
                        \hline
                    \end{tabular}
                }
            };
            \node at (-16,-20){
                \scalebox{3.5}{
                    \begin{tabular}{|c|c|}
                        \hline
                        \textbf{Destination} & \textbf{Next Hop} \\ 
                        \hline
                        Default Route & fe80::8aa:ff:fe00:2\\  
                        \hline
                    \end{tabular}
                }
            };
            \node at (4,-20){
                \scalebox{3.5}{
                    \begin{tabular}{|c|c|}
                        \hline
                        \textbf{Destination} & \textbf{Next Hop} \\ 
                        \hline
                        Default Route & fe80::8aa:ff:fe00:2\\  
                        \hline
                    \end{tabular}
                }
            };
						 \node at (16,-14){
                \scalebox{3.5}{
                    \begin{tabular}{|c|c|}
                        \hline
                        \textbf{Destination} & \textbf{Next Hop} \\ 
                        \hline
                        Default Route & fe80::8aa:ff:fe00:3\\  
                        \hline
                    \end{tabular}
                }
            };
						 \node at (20,-27){
                \scalebox{3.5}{
                    \begin{tabular}{|c|c|}
                        \hline
                        \textbf{Destination} & \textbf{Next Hop} \\ 
                        \hline
                        Default Route & fe80::8aa:ff:fe00:6\\  
                        \hline
                    \end{tabular}
                }
            };

            \begin{scope}[>={Stealth[black]},
                every edge/.style={draw=black,very thick}]
                \path [-] (2) edge node {} (1);
                \path [-] (3) edge node {} (1);
                \path [-] (4) edge node {} (2);
                \path [-] (5) edge node {} (2);
                \path [-] (6) edge node {} (3);
                \path [-] (7) edge node {} (6);
            \end{scope}
        \end{tikzpicture}
    \end{adjustbox}
    \caption{Upward Routes in~\ac{mop} 0}
    \label{fig:upwardRoute:mop0}
\end{figure}
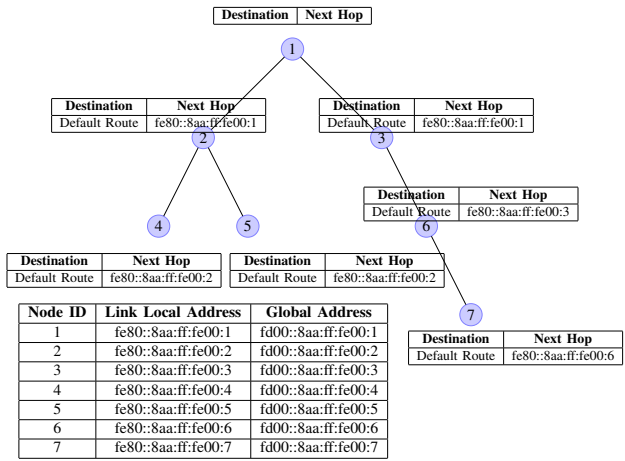

In this \ac{mop}, all nodes excluding a \ac{dodag} root node maintain a routing table to store a default route to specify a next hop to reach the \ac{dodag} root (Fig.~\ref{fig:upwardRoute:mop0}). Upward routes and \ac{dodag} are constructed and maintained by propagating \ac{dio} messages. When a child node receives a \ac{dio} message, the child node selects the sender of the \ac{dio} message as a parent node and creates a default route to the parent using the link local address of the parent in the routing table if the \ac{dio} message passes the rules specified in the standard such as having a valid version number, rank, \ac{dodag} ID and etc. After constructing the Upward routes, if a child node wants to send a data packet to a \ac{dodag} root, it routes the packet to a default route specifing the \ac{dodag} root.

\subsection{MOPs 1 and 2}
Both \ac{mop} 1 and 2 use Downward routes in addition to the Upward routes. Using Downward routes enables the protocol to route \ac{p2mp} traffic type, that is, to route data packets downward from a \ac{dodag} root to other child nodes along the \ac{dodag}. Since these \ac{mop}s can route using both Upward and Downward routes, they can route \ac{p2p} traffic type along the \ac{dodag} in different ways. Downward routes are established and maintained by propagating \ac{dao} messages upward along the \ac{dodag}.

\begin{figure}[hbt]
    \centering
    \begin{adjustbox}{width=0.48\textwidth}
        \tikzstyle{myNetworkNode}=[circle,draw=blue!50,fill=blue!20,thick,inner sep=0pt,minimum size=20mm]
        \tikzstyle{myEdge}=[]
        \tikzstyle{myCalloutUp}=[ellipse callout, fill=blue!20,opacity=.8, yshift=1.5cm]
        \tikzstyle{myCalloutDown}=[ellipse callout, fill=blue!20,opacity=.8, yshift=-1.5cm]
        \tikzstyle{myAddressTable}=[]
        \begin{tikzpicture}
            \node[myNetworkNode] (1) at (0,0) {\scalebox{4}{1}};
            \node[myNetworkNode] (2) at (-8,-8) {\scalebox{4}{2}};
            \node[myNetworkNode] (3) at (8,-8) {\scalebox{4}{3}};
            \node[myNetworkNode] (4) at (-12,-16) {\scalebox{4}{4}};
            \node[myNetworkNode] (5) at (-4,-16) {\scalebox{4}{5}};
            \node[myNetworkNode] (6) at (12,-16) {\scalebox{4}{6}};
            \node[myNetworkNode] (7) at (16,-24) {\scalebox{4}{7}};

            \node[myAddressTable] at (-8,-30) {
                \scalebox{4}{                
                    \begin{tabular}{|c|c|c|}
                        \hline
                        \textbf{Node ID} & \textbf{Link Local Address} & \textbf{Global Address} \\ 
                        \hline
                        1 & fe80::8aa:ff:fe00:1 & fd00::8aa:ff:fe00:1 \\ 
                        \hline
                        2 & fe80::8aa:ff:fe00:2 & fd00::8aa:ff:fe00:2 \\ 
                        \hline
                        3 & fe80::8aa:ff:fe00:3 & fd00::8aa:ff:fe00:3 \\ 
                        \hline
                        4 & fe80::8aa:ff:fe00:4 & fd00::8aa:ff:fe00:4 \\ 
                        \hline
                        5 & fe80::8aa:ff:fe00:5 & fd00::8aa:ff:fe00:5 \\ 
                        \hline
                        6 & fe80::8aa:ff:fe00:6 & fd00::8aa:ff:fe00:6 \\ 
                        \hline
                        7 & fe80::8aa:ff:fe00:7 & fd00::8aa:ff:fe00:7 \\ 
                        \hline
                    \end{tabular}
                }
            };

            \node at (0,7){
                \scalebox{3.5}{
                    \begin{tabular}{|c|c|}
                        \hline
                        \textbf{Child} & \textbf{DAO Parent} \\ 
												\hline
												fd00::8aa:ff:fe00:1 & NULL \\
                        \hline
												fd00::8aa:ff:fe00:2 & fd00::8aa:ff:fe00:1 \\ 
                        \hline
                        fd00::8aa:ff:fe00:3 & fd00::8aa:ff:fe00:1 \\ 
                        \hline
                        fd00::8aa:ff:fe00:4 & fd00::8aa:ff:fe00:2\\ 
                        \hline
                        fd00::8aa:ff:fe00:5 & fd00::8aa:ff:fe00:2\\ 
                        \hline
                        fd00::8aa:ff:fe00:6 & fd00::8aa:ff:fe00:3\\ 
                        \hline
                        fd00::8aa:ff:fe00:7 & fd00::8aa:ff:fe00:6\\ 
                        \hline											
                    \end{tabular}
                }
            };
            \node at (-12,-6){
                \scalebox{3.5}{
                    \begin{tabular}{|c|c|}
                        \hline
                        \textbf{Destination} & \textbf{Next Hop} \\ 
                        \hline
                        Default Route & fe80::8aa:ff:fe00:1\\  
                        \hline
                    \end{tabular}
                }
            };
            \node at (12,-6){
                \scalebox{3.5}{
                    \begin{tabular}{|c|c|}
                        \hline
                        \textbf{Destination} & \textbf{Next Hop} \\ 
                        \hline
                        Default Route & fe80::8aa:ff:fe00:1\\  
                        \hline
                    \end{tabular}
                }
            };
            \node at (-16,-20){
                \scalebox{3.5}{
                    \begin{tabular}{|c|c|}
                        \hline
                        \textbf{Destination} & \textbf{Next Hop} \\ 
                        \hline
                        Default Route & fe80::8aa:ff:fe00:2\\  
                        \hline
                    \end{tabular}
                }
            };
            \node at (4,-20){
                \scalebox{3.5}{
                    \begin{tabular}{|c|c|}
                        \hline
                        \textbf{Destination} & \textbf{Next Hop} \\ 
                        \hline
                        Default Route & fe80::8aa:ff:fe00:2\\  
                        \hline
                    \end{tabular}
                }
            };
						 \node at (16,-14){
                \scalebox{3.5}{
                    \begin{tabular}{|c|c|}
                        \hline
                        \textbf{Destination} & \textbf{Next Hop} \\ 
                        \hline
                        Default Route & fe80::8aa:ff:fe00:3\\  
                        \hline
                    \end{tabular}
                }
            };
						 \node at (20,-27){
                \scalebox{3.5}{
                    \begin{tabular}{|c|c|}
                        \hline
                        \textbf{Destination} & \textbf{Next Hop} \\ 
                        \hline
                        Default Route & fe80::8aa:ff:fe00:6\\  
                        \hline
                    \end{tabular}
                }
            };

            \begin{scope}[>={Stealth[black]},
                every edge/.style={draw=black,very thick}]
                \path [-] (2) edge node {} (1);
                \path [-] (3) edge node {} (1);
                \path [-] (4) edge node {} (2);
                \path [-] (5) edge node {} (2);
                \path [-] (6) edge node {} (3);
                \path [-] (7) edge node {} (6);
            \end{scope}
        \end{tikzpicture}
    \end{adjustbox}
    \caption{Upward/Downward Routes in~\ac{mop} 1}
    \label{fig:upDownwardRoute:mop1}
\end{figure}
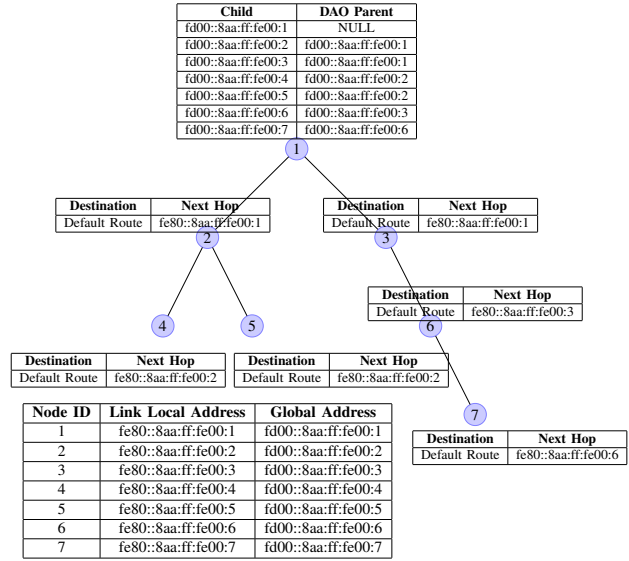

More specifically, \ac{mop} 1 applies source routing to route data packets downward. As shown in Fig.~\ref{fig:upDownwardRoute:mop1}, \ac{dodag} root node only maintains a table, called source routing table, and other child nodes do not store any routing entries to route packets downward, hence this \ac{mop} is also known as \textbf{Non-Storing Mode of Operation}. 
To establish routes, \ac{dao} messages are sent as unicast messages by child nodes to the \ac{dodag} root node. Since the destination of the message is the \ac{dodag} root, the \ac{dodag} is routed upward by the default route entries using Upward routes like a data packet of the \ac{mp2p} traffic type. When the \ac{dodag} root receives the message, it creates an entry in its source routing table. After establishing Downward routes, when a \ac{dodag} root wants to send a data packet to a child node, the \ac{dodag} root recursively generates a sequence of intermediate nodes from itself to the child node to be accommodated in the source routing header since the \ac{dodag} root has all routing information in its source routing table. As shown in Fig.~\ref{fig:upDownwardRoute:mop1}, both child and selected parent are global addresses in the source routing table.

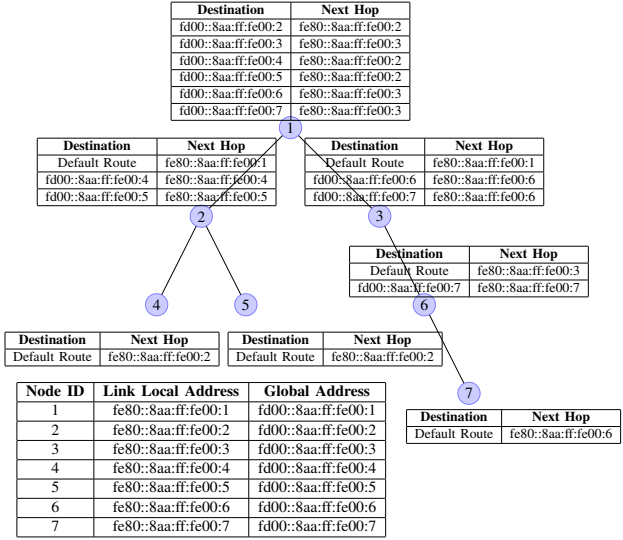
\begin{figure}[hbt]
    \centering
    \begin{adjustbox}{width=0.48\textwidth}
        \tikzstyle{myNetworkNode}=[circle,draw=blue!50,fill=blue!20,thick,inner sep=0pt,minimum size=20mm]
        \tikzstyle{myEdge}=[]
        \tikzstyle{myCalloutUp}=[ellipse callout, fill=blue!20,opacity=.8, yshift=1.5cm]
        \tikzstyle{myCalloutDown}=[ellipse callout, fill=blue!20,opacity=.8, yshift=-1.5cm]
        \tikzstyle{myAddressTable}=[]
        \begin{tikzpicture}
            \node[myNetworkNode] (1) at (0,0) {\scalebox{4}{1}};
            \node[myNetworkNode] (2) at (-8,-8) {\scalebox{4}{2}};
            \node[myNetworkNode] (3) at (8,-8) {\scalebox{4}{3}};
            \node[myNetworkNode] (4) at (-12,-16) {\scalebox{4}{4}};
            \node[myNetworkNode] (5) at (-4,-16) {\scalebox{4}{5}};
            \node[myNetworkNode] (6) at (12,-16) {\scalebox{4}{6}};
            \node[myNetworkNode] (7) at (16,-24) {\scalebox{4}{7}};

            \node[myAddressTable] at (-8,-30) {
                \scalebox{4}{                
                    \begin{tabular}{|c|c|c|}
                        \hline
                        \textbf{Node ID} & \textbf{Link Local Address} & \textbf{Global Address} \\ 
                        \hline
                        1 & fe80::8aa:ff:fe00:1 & fd00::8aa:ff:fe00:1 \\ 
                        \hline
                        2 & fe80::8aa:ff:fe00:2 & fd00::8aa:ff:fe00:2 \\ 
                        \hline
                        3 & fe80::8aa:ff:fe00:3 & fd00::8aa:ff:fe00:3 \\ 
                        \hline
                        4 & fe80::8aa:ff:fe00:4 & fd00::8aa:ff:fe00:4 \\ 
                        \hline
                        5 & fe80::8aa:ff:fe00:5 & fd00::8aa:ff:fe00:5 \\ 
                        \hline
                        6 & fe80::8aa:ff:fe00:6 & fd00::8aa:ff:fe00:6 \\ 
                        \hline
                        7 & fe80::8aa:ff:fe00:7 & fd00::8aa:ff:fe00:7 \\ 
                        \hline
                    \end{tabular}
                }
            };

            \node at (0,6){
                \scalebox{3.5}{
                    \begin{tabular}{|c|c|}
                        \hline
                        \textbf{Destination} & \textbf{Next Hop} \\ 
                        \hline
                        fd00::8aa:ff:fe00:2 & fe80::8aa:ff:fe00:2 \\ 
                        \hline
                        fd00::8aa:ff:fe00:3 & fe80::8aa:ff:fe00:3 \\ 
                        \hline
                        fd00::8aa:ff:fe00:4 & fe80::8aa:ff:fe00:2\\ 
                        \hline
                        fd00::8aa:ff:fe00:5 & fe80::8aa:ff:fe00:2\\ 
                        \hline
                        fd00::8aa:ff:fe00:6 & fe80::8aa:ff:fe00:3\\ 
                        \hline
                        fd00::8aa:ff:fe00:7 & fe80::8aa:ff:fe00:3\\ 
                        \hline											
                    \end{tabular}
                }
            };
            \node at (-12,-4){
                \scalebox{3.5}{
                    \begin{tabular}{|c|c|}
                        \hline
                        \textbf{Destination} & \textbf{Next Hop} \\ 
                        \hline
                        Default Route & fe80::8aa:ff:fe00:1\\  
                        \hline
												fd00::8aa:ff:fe00:4 & fe80::8aa:ff:fe00:4\\ 
                        \hline
                        fd00::8aa:ff:fe00:5 & fe80::8aa:ff:fe00:5\\ 
                        \hline
                    \end{tabular}
                }
            };
            \node at (12,-4){
                \scalebox{3.5}{
                    \begin{tabular}{|c|c|}
                        \hline
                        \textbf{Destination} & \textbf{Next Hop} \\ 
                        \hline
                        Default Route & fe80::8aa:ff:fe00:1\\  
                        \hline
												fd00::8aa:ff:fe00:6 & fe80::8aa:ff:fe00:6\\ 
                        \hline
                        fd00::8aa:ff:fe00:7 & fe80::8aa:ff:fe00:6\\ 
                        \hline	
                    \end{tabular}
                }
            };
            \node at (-16,-20){
                \scalebox{3.5}{
                    \begin{tabular}{|c|c|}
                        \hline
                        \textbf{Destination} & \textbf{Next Hop} \\ 
                        \hline
                        Default Route & fe80::8aa:ff:fe00:2\\  
                        \hline
                    \end{tabular}
                }
            };
            \node at (4,-20){
                \scalebox{3.5}{
                    \begin{tabular}{|c|c|}
                        \hline
                        \textbf{Destination} & \textbf{Next Hop} \\ 
                        \hline
                        Default Route & fe80::8aa:ff:fe00:2\\  
                        \hline
                    \end{tabular}
                }
            };
						 \node at (16,-13){
                \scalebox{3.5}{
                    \begin{tabular}{|c|c|}
                        \hline
                        \textbf{Destination} & \textbf{Next Hop} \\ 
                        \hline
                        Default Route & fe80::8aa:ff:fe00:3\\  
                        \hline
												fd00::8aa:ff:fe00:7 & fe80::8aa:ff:fe00:7\\ 
                        \hline
                    \end{tabular}
                }
            };
						 \node at (20,-27){
                \scalebox{3.5}{
                    \begin{tabular}{|c|c|}
                        \hline
                        \textbf{Destination} & \textbf{Next Hop} \\ 
                        \hline
                        Default Route & fe80::8aa:ff:fe00:6\\  
                        \hline
                    \end{tabular}
                }
            };

            \begin{scope}[>={Stealth[black]},
                every edge/.style={draw=black,very thick}]
                \path [-] (2) edge node {} (1);
                \path [-] (3) edge node {} (1);
                \path [-] (4) edge node {} (2);
                \path [-] (5) edge node {} (2);
                \path [-] (6) edge node {} (3);
                \path [-] (7) edge node {} (6);
            \end{scope}
        \end{tikzpicture}   
    \end{adjustbox}
    \caption{Upward/Downward Routes in~\ac{mop} 2}
    \label{fig:upDownwardRoute:mop2}
\end{figure}

In the case of \ac{mop} 2, it applies the traditional hop-by-hop routing maintaining traffic flow state information at each intermediate node to route packets downward. 
As depicted in Fig.~\ref{fig:upDownwardRoute:mop2}, all nodes maintain a routing table to store routing entries of their sub-tree, hence this \ac{mop} is described as \textbf{Storing Mode of Operation}. 
\ac{dao} messages are sent as unicast messages by child nodes to a selected parent node(s). When a \ac{dao} message is sent to a selected parent node, the parent node creates an entry in its routing table. We encounter the concepts of the forwarder and generator nodes in this \ac{mop}. The generator node creates the \ac{dao} message and puts its global address as the destination advertised address field in the message and inserts its link local address as the sender address in the packet, then it sends the message to the selected parent node. The parent node creates an entry in its routing table after receiving the message. The entry includes the sender's address as the next hop field and the destination advertised address as the destination field. Briefly, the parent node updates the sender address of the packet to its link local address, then it forwards the message to its selected parent. This process continues until the message reaches the \ac{dodag} root.

\section{Network simulators and emulators}
\label{platforms}

In order to implement RPL to evaluate it and compare it with other enhanced or new competing protocols, we first assessed the different network simulators and emulators, commonly leveraged in the field for this purpose.

In this sense, the most popular platform for \ac{iot} environments is \textbf{Contiki-NG}~\cite{contiki-ng}. Contiki-NG is an open-source operating system implementing the standard protocol stack containing IPv6/6LoWPAN, 6TiSCH, \ac{rpl}, and CoAP for next-generation \ac{iot} devices and low-power communication. 
Contiki-NG includes real platforms such as the Sky mote~\cite{sky}. Because of RAM and ROM limitations, the Sky mote cannot run implementations which have a lot of code. Furthermore, Contiki-NG includes a virtual platform, called Cooja mote, which is a fast mote not constrained in memory, although simulation with Cooja mote is not perfectly time-accurate. In addition to compile and load implemented code in a real device, Contiki-NG includes a simulator, called Cooja~\cite{cooja}, which can emulate real platforms and simulate virtual platforms to run and test the implemented code. 
Additionally, motes programmed with Contiki-NG can be easily deployed in large \ac{iot} environments and testbeds like FIT IoT-LAB~\cite{iotlab}, which increases

Another two well-known simulators are \textbf{ns-3}~\cite{ns3} and \textbf{OMNeT++}~\cite{omnetpp}. 
On the one hand, ns-3 is a discrete-event network simulator for Internet systems and \ac{rpl} is available for this simulator too~\cite{bartolozzi12}. 
On the other hand, OMNeT++~\cite{omnetpp} is a discrete-event simulator in which several frameworks (such as INET~\cite{inet}, MiXiM~\cite{mixim}) can be executed. Kermajani \textit{et al.}~\cite{kermajani14} implemented a preliminary version of \ac{rpl}. This implementation only supports \ac{mop} 0 to construct Upward routes. 
In this regard, both of them are packet-based simulators and we could say --in a very simplistic way-- that OMNeT++ is usually easier to program as it provides a higher-level view, while ns-3 has a better overall performance.

Additionally, there is a third library for simulation that is worth mentioning, in this case event-based, which is called \textbf{SimPy}~\cite{simpy}. As the name indicates, it is based on Python and allos the implementation of high-level simulations (e.g., flow-based) in a fast and easy way.

Up to this point, all described platforms are used to implement different protocols in a distributed manner, i.e., the logic of the protocol is implemented as the logic of a network node, which is later on deployed in a network scenario. 
However, there is an alternative centralized approach that can be followed by applying the \ac{sdn}~\cite{sdn-survey} paradigm, which is not only an architecture, but also another mean of testing and validation of protocols. 
Although \ac{sdn} is leveraged in many implementations for realistic developments based on a centralized logic developed in the form of a software, most of the current platforms are mainly focused on non-constrained wired network devices, which makes it difficult to apply in \acp{lln}. 
There is a current effort to implement a platform to test these networks called Mininet-WiFi~\cite{mininet-wifi}, which is a extension of Mininet~\cite{mininet}, and also some frameworks like SDN-WISE~\cite{Galluccio15} try to integrate both Contiki-NG motes and \ac{sdn}. 

\subsection{Reason to select OMNeT++ as implementation platform}

In our case, we wanted to test a completely new protocol to be compared with RPL. As the protocol was designed from scratch and not as an extension of RPL, we believed the first step was to validate it with a simple and fast prototype. 
In this regard, although the multiple benefits of Contiki-NG are apparent, this environment might be slightly harsh for beginners, and the implied effort is high if we consider we just want to validate an initial idea. 
For this reason, we thought OMNeT++ was the simplest way for this initial step, because it was simpler than ns-3 and Contiki-NG, while keeping more realistic than SimPy (which is not packet-based). 

For the implementation of RPL we took the work from Kermajani \textit{et al.}~\cite{kermajani14} as a reference. 
Since the MiXiM framework was not updated because it is not supported by the new versions of OMNeT++, we implemented \ac{rpl} in OMNeT++ 5.2.1 and the INET 3.6.3 framework~\cite{inet363} instead. This implementation had to support other \ac{mop}s applying the Downward routes, ICMPv6 messages for transmitting the RPL control messages, the interface and routing table. 
In particular, our implementation followed the standard~\cite{rfc6550}, the implementation of \ac{rpl} in MiXiM~\cite{kermajani14}, and Contiki-NG~\cite{contiki-ng}. Unlike Contiki-NG, the simulation runs a single instance and \ac{dodag} of \ac{rpl}.

\section{Implementation}
\label{implementation}

Since our objective was to fully implement the routing functionality of RPL, according to the definition of RPL provided in Section~\ref{rpl}, we implemented the three \acp{mop}: 0, 1 and 2. 
In particular, we focused on the network layer, while the link and physical layers were kept simple (though extensible in the future if required). 

As a departing point, we considered the original INET framework and we modified it accordingly. In the following paragraphs, we explain the applied changes and the expected (and resulting) behavior. 

\begin{figure}
        \centering
        \begin{subfigure}[htb]{0.48\textwidth}
                \includegraphics[width=\textwidth]{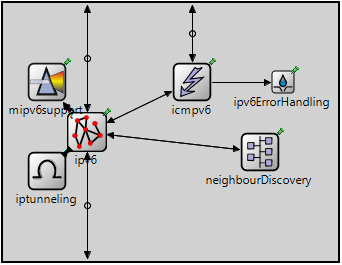}
                \caption{Original design of the \texttt{Network Layer} in INET 3.6.3}
                \label{fig:netLayerINET}
        \end{subfigure}%
        \quad
        \begin{subfigure}[htb]{0.48\textwidth}
                \includegraphics[width=\textwidth]{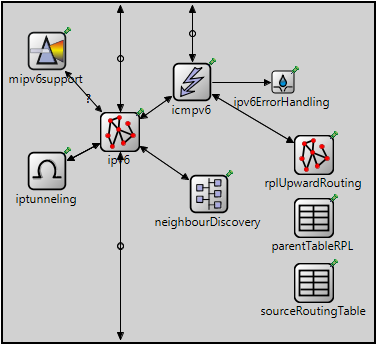}
                \caption{Extended design of the \texttt{Network Layer} (after implementing RPL)}
                \label{fig:netLayerRPL}
        \end{subfigure}
        \caption{Comparison design of the \texttt{Network Layer}}
				\label{fig:netLayer}
\end{figure}

Fig.~\ref{fig:netLayer} illustrate an outline of the original \texttt{Network Layer} module defined in INET (Fig.~\ref{fig:netLayerINET}) compared to the implemented extension of it, which comprises the definition of RPL (Fig.~\ref{fig:netLayerRPL}). 
We added three submodules to module \texttt{IPv6NetworkLayer}: (1) \texttt{ParentTableRPL} maintains parent node information, (2) \texttt{sourceRoutingTable} maintains source routing entries in \ac{mop} 1, and finally, (3) \texttt{RPLUpwardRouting} is a submodule to implement the routing functionality of Upward routes for all \ac{mop}s. 
Since the upward routing functionality was more complex than the downward one and, at the same time, to decrease the complexity of the \texttt{ICMPv6} submodule, we separated the functionality from this \texttt{ICMPv6} module. In addition to the newly added submodules, we also updated some existing modules of INET shown in Table~\ref{table:updatedModules}. 

\begin{table*}[!h]
	\centering
	\caption{Updated modules/classes in the implementation}
	\label{table:updatedModules}
	\begin{tabular}{|p{0.3\textwidth}|p{0.3\textwidth}|p{0.35\textwidth}|}
		\hline
		\textbf{Module} & \textbf{Updated module} & \textbf{Functionality description}\\
		\hline
		\texttt{ICMPv6} & \texttt{ICMPv6RPL} & To implement the Downward routing for \ac{mop} 1 and 2\\
		\hline
		\texttt{ICMPv6Message} & \texttt{ICMPv6MessageRPL} & To define the ICMPv6 \ac{rpl} control messages\\
		\hline
		\texttt{IPv6NeighbourDiscoveryRPL} & \texttt{IPv6NeighbourDiscoveryRPL} & To maintain neighbors discoverd by the \ac{rpl} control messages\\
		\hline
		\texttt{IPv6NeighbourCache} & \texttt{IPv6NeighbourCacheRPL} & To maintain neighbors discovered by the \ac{rpl} control messages\\
		\hline
		\texttt{IPv6} & \texttt{IPv6RPL} & To handle the ICMPv6 \ac{rpl} control messages\\
		\hline
	\end{tabular}
\end{table*}

As mentioned in the previous section, we use both link local and global addresses to form the \ac{dodag}, Upward and Downward routes. 
To assign addresses to each node, we updated the \texttt{IPv6NeighbourDiscovery} module in INET to \texttt{IPv6NeighbourDiscoveryRPL} in our implementation. 
One of the updated changes is to statically assign a link local address to each node. In order to assign a global address to each node, we first check whether a global address has been assigned to the node or not. If there is not any assigned global address to the node, we assign a global address to the node. The prefix of the global address is \texttt{fd00::/64}, which is the prefix used in Contiki-NG, and its suffix is a sequence of last 64 bits of the link local address. 
Since the functionality of the global address assignment is the same function in the three simulated~\ac{mop}s of~\ac{rpl}, we placed the function in the \texttt{RPLUpwardRouting} module.

\subsection{Message propagation and processing (\texttt{Network} layer)}

\subsubsection{\ac{dio}}

First, the \ac{dodag} root node propagates a \ac{dio} message. Then, other nodes which receive the message, update the message and schedule a timer to send it if some situations are satisfied (e.g., avoiding a loop, receiving invalid messages by checking the version, etc.). Submodule \texttt{rplUpwardRouting} performs the operation.

When the \texttt{Network Layer} receives a \ac{dio} message from the \texttt{MAC Layer}, the message is first sent to submodule \texttt{ipv6}. In the submodule, the message is duplicated. Then, a copy of the message is sent to submodule \texttt{neighbourDiscovery} to update the submodule's entries based on rules, and the original message is sent to submodule \texttt{icmpv6}. Since the \ac{dio} messages are used for constructing Upward routes, submodule \texttt{icmpv6} sends the message to submodule \texttt{rplUpwardRouting}, where needed processes are done (e.g., if needed to update \texttt{parentTableRPL} or not, adding a default route to module \texttt{routingTable} or not, etc.). These operations are done for all \ac{mop}s.

\subsubsection{\ac{dis}}

If non-root nodes have not joined to a \ac{dodag} after some specified time, submodule \texttt{icmpv6} schedules a \ac{dis} message to be sent.

Like the received \ac{dio} message, when the \texttt{Network Layer} receives a \ac{dis} message from the \texttt{MAC Layer}, the original message is sent to submodule \texttt{icmpv6}, and a duplicated copy is sent to submodule \texttt{neighbourDiscovery}. Unlike the received \ac{dio} message, submodule \texttt{icmpv6} processes the \ac{dis} message since the process is not complex. 

\subsubsection{\ac{dao}}

After receiving a \ac{dio} message which introduces a new parent node, a \ac{dao} message is scheduled to be sent.

When the \texttt{Network Layer} receives a \ac{dao} message from the \texttt{MAC Layer}, different actions are performed in each \ac{mop}. 
In \ac{mop} 0, no routing operations are performed for Upward routes, so no \ac{dao} messages are propagated in \ac{mop} 0. 
In \ac{mop} 1, the message is routed to the \ac{dodag} root node by the default route dedicated to the \ac{dodag} root, so it is sent to the parent node. Simultaneously, submodule \texttt{ipv6} sends a copy of the message to submodule \texttt{neighbourDiscovery}. Finally, the \ac{dodag} root node creates an entry in the source routing table based on the message's originator, as the child node, and the parent of the originator, as the \ac{dao} parent node when receiving the message. 
In \ac{mop} 2, a duplicated copy is sent to submodule \texttt{neighbourDiscovery} and the original one is sent to submodule \texttt{icmpv6}. Like a \ac{dis} message, submodule \texttt{icmpv6} processes the \ac{dao} message to update/add an entry in the routing table and sends a \ac{dao} message to a parent node sometimes when some conditions are satisfied. 




\subsection{\texttt{MAC} and \texttt{Physical} layers}

We designed a simple ideal \texttt{MAC} and \texttt{Physical} layers to check \ac{rpl} functionality in ideal conditions. \texttt{IEEE802.15.4} is easily replaceable with the ideal \texttt{MAC Layer} by putting \texttt{**.wlanType = "Ieee802154NarrowbandNic"} in the \texttt{.ini} ffile of an OMNeT++ project.
\section{How to install and use the implementation}
\label{installation}

The implemented code is openly available in GitHub~\cite{simulation}. 
To install it, the next steps should be followed:
\begin{itemize}
	\item Install OMNeT++
	\begin{itemize}
		\item Download OMNeT++ 5.2.1~\cite{omnetpp}.
		\item Let OMNeT++ install and use its compiler. After extracting and before installing OMNeT++, change "PREFER\_CLANG=yes" to "PREFER\_CLANG=no" in the "configure.user" file in your OMNeT++ installation folder.
		\item Install OMNeT++.
		\item To check whether OMNeT++ correctly works or not, run an example of OMNeT++ such as dyna, aloha, tictoc, etc.
	\end{itemize}
	\item Install INET
	\begin{itemize}
		\item Download INET 3.6.3~\cite{inet363}.
		\item Install and build INET.
		\item To check whether INET correctly works, run an example of INET such as inet/examples/adhoc/ieee80211, etc.
	\end{itemize}
	\item Install the RPL implementation
	\begin{itemize}
		\item Since the simulation files only include implementation code, the project may not be probably imported by the IDE, and a new OMNeT++ project must be manually created. Therefore, please follow the next substeps:
		\begin{itemize}
			\item Click on the menu of "File".
			\item Select "New".
			\item Select "OMNeT++ Project".
			\item Type the name of project.
			\item Click on "Next".
			\item Select "Empty Project".
			\item Click on "Finish".
			\item Copy all files/folders in the folder to the created project folder.
		\end{itemize}
		\item Introduce INET to the project as a reference/library, Therefore, please follow the next substeps:
		\begin{itemize}
			\item Right-click on the project in "Project Explorer" window.
			\item Select "Properties".
			\item Select "Project Reference" in the left list.
			\item Select "INET" in the right list.
			\item Click on the "Apply and close".
		\end{itemize}
		\item Build and run the project.
	\end{itemize}
\end{itemize}
\section{Results, Expected Impact and Contributions}
\label{results}

The implementation of RPL in OMNeT++ allowed our team to quickly evaluate and compare this protocol with our own proposal, entitled IoTorii~\cite{Rojas21}. We created scenarios of up to 50 motes to have initial results regarding different parameters such as number of table entries and convergence time, which helped to polish the design of our protocol, which was eventually implemented in Contiki-NG, obtaining very similar results. Therefore, this implementation helped immensely to prove the initial design and analysis of our protocol, in a fast and easy way. 
However, the main shortcoming is related with the physical layer, which is currently ideal and does not reflect completely realistic scenarios in that regard. 

As previously mentioned, the code is publicly available in GitHub~\cite{simulation} and some other researchers are already using and testing it, as it can be observed in their public forums. 
The simulations support more than 200 motes, which can be useful for big scenarios, before testing them with Contiki-NG, which might required more computational resources. 
For this reason, we expect some impact in the upcoming years from the usage of our implementation.

However, the OMNeT++ and INET frameworks evolved rapidly, and it is difficult to keep all versions up to date. Although the current version of the code is still usable, we kindly invite any researcher to clone the code and update it or extend it, as any contribution would be probably highly appreciated by the research community.
\section{Conclusion and Future Work}
\label{conclusion}

In this article, we have presented the background and design decision of our implementation of the RPL protocol in the OMNeT++ simulator. 
Although our research team has experience in all the beforementioned platforms (Contiki-NG, ns-3, SimPy, and SDN-related platforms like Mininet or the Ryu and ONOS controller), our conclusion is that OMNeT++ is an ideal first step to test very conceptual research approaches (particularly if based in in designs from scratch) and for beginners in the field. For that reason, we have made the code publicly available. 

As future work, we would like to extend the implementation to include a realistic physical layer, as this is particularly relevant in \acp{lln}. 
Moreover, we would like to invite any researcher in the field to use and contribute to our code if willing to.

\section*{Acknowledgment}
This work was funded by grants from Comunidad de Madrid through project TAPIR-CM (S2018/TCS-4496) and project IRIS-CM (CM/JIN/2019-039), by Junta de Comunidades de Castilla-La Mancha through project IRIS-JCCM (SBPLY/19/180501/000324), and by Universidad de Alcal\'a through project DEDENNE (CCG20/IA-012). 

Additionally, we would also like to thank Carles Gomez for his help and guidance during the implementation process.


\ifCLASSOPTIONcaptionsoff
  \newpage
\fi

\bibliographystyle{IEEEtran}
\bibliography{RPL}   

\begin{thebibliography}{10}
\providecommand{\url}[1]{#1}
\csname url@samestyle\endcsname
\providecommand{\newblock}{\relax}
\providecommand{\bibinfo}[2]{#2}
\providecommand{\BIBentrySTDinterwordspacing}{\spaceskip=0pt\relax}
\providecommand{\BIBentryALTinterwordstretchfactor}{4}
\providecommand{\BIBentryALTinterwordspacing}{\spaceskip=\fontdimen2\font plus
\BIBentryALTinterwordstretchfactor\fontdimen3\font minus
  \fontdimen4\font\relax}
\providecommand{\BIBforeignlanguage}[2]{{%
\expandafter\ifx\csname l@#1\endcsname\relax
\typeout{** WARNING: IEEEtran.bst: No hyphenation pattern has been}%
\typeout{** loaded for the language `#1'. Using the pattern for}%
\typeout{** the default language instead.}%
\else
\language=\csname l@#1\endcsname
\fi
#2}}
\providecommand{\BIBdecl}{\relax}
\BIBdecl

\bibitem{Tomkos20}
I.~{Tomkos}, D.~{Klonidis}, E.~{Pikasis}, and S.~{Theodoridis}, ``{Toward the
  6G Network Era: Opportunities and Challenges},'' \emph{IT Professional},
  vol.~22, no.~1, pp. 34--38, 2020.

\bibitem{Carrascal20}
\BIBentryALTinterwordspacing
D.~Carrascal, E.~Rojas, J.~Alvarez-Horcajo, D.~Lopez-Pajares, and
  I.~Mart\'inez-Yelmo, ``{Analysis of P4 and XDP for IoT Programmability in 6G
  and Beyond},'' \emph{IoT}, vol.~1, no.~2, p. 605–622, Dec 2020. [Online].
  Available: \url{http://dx.doi.org/10.3390/iot1020031}
\BIBentrySTDinterwordspacing

\bibitem{rfc6550}
T.~Winter, P.~Thubert, A.~Brandt, J.~Hui, R.~Kelsey, P.~Levis, K.~Pister,
  R.~Struik, J.~P. Vasseur, and R.~Alexander, ``{RFC 6550: RPL: IPv6 Routing
  Protocol for Low-Power and Lossy Networks (2012)},'' \emph{URL https://tools.
  ietf. org/html/rfc6550}, 2012.

\bibitem{contiki-ng}
\BIBentryALTinterwordspacing
``{Contiki-NG: The OS for Next Generation IoT Devices}.'' [Online]. Available:
  \url{http://www.contiki-ng.org/}
\BIBentrySTDinterwordspacing

\bibitem{sky}
\BIBentryALTinterwordspacing
``{sky mote}.'' [Online]. Available:
  \url{https://github.com/contiki-ng/contiki-ng/wiki/Platform-sky}
\BIBentrySTDinterwordspacing

\bibitem{cooja}
A.~Velinov and A.~Mileva, ``{Running and testing applications for Contiki OS
  using Cooja simulator},'' 2016.

\bibitem{iotlab}
\BIBentryALTinterwordspacing
``{FIT IoT-LAB: The Very Large Scale IoT Testbed}.'' [Online]. Available:
  \url{https://www.iot-lab.info/}
\BIBentrySTDinterwordspacing

\bibitem{ns3}
\BIBentryALTinterwordspacing
``{ns-3 Network Simulator}.'' [Online]. Available: \url{https://www.nsnam.org/}
\BIBentrySTDinterwordspacing

\bibitem{omnetpp}
\BIBentryALTinterwordspacing
``{OMNeT++ Simulator}.'' [Online]. Available: \url{https://omnetpp.org/}
\BIBentrySTDinterwordspacing

\bibitem{bartolozzi12}
L.~Bartolozzi, T.~Pecorella, and R.~Fantacci, ``{ns-3 RPL module: IPv6 routing
  protocol for low power and lossy networks},'' in \emph{Proceedings of the 5th
  international ICST conference on simulation tools and techniques}.\hskip 1em
  plus 0.5em minus 0.4em\relax ICST (Institute for Computer Sciences,
  Social-Informatics and Telecommunications Engineering), 2012, pp. 359--366.

\bibitem{inet}
\BIBentryALTinterwordspacing
``{INET Framework}.'' [Online]. Available: \url{https://inet.omnetpp.org/}
\BIBentrySTDinterwordspacing

\bibitem{mixim}
\BIBentryALTinterwordspacing
``{MiXiM Framework}.'' [Online]. Available:
  \url{http://mixim.sourceforge.net/index.html}
\BIBentrySTDinterwordspacing

\bibitem{kermajani14}
H.~Kermajani and C.~Gomez, ``{On the network convergence process in RPL over
  IEEE 802.15. 4 multihop networks: Improvement and trade-offs},''
  \emph{Sensors}, vol.~14, no.~7, pp. 11\,993--12\,022, 2014.

\bibitem{simpy}
N.~Matloff, ``{Introduction to Discrete-Event Simulation and the SimPy
  Language},'' \emph{Davis, CA. Dept of Computer Science. University of
  California at Davis. Retrieved on August}, vol.~2, no. 2009, pp. 1--33, 2008.

\bibitem{sdn-survey}
D.~Kreutz, F.~M.~V. Ramos, P.~E. Ver\'issimo, C.~E. Rothenberg, S.~Azodolmolky,
  and S.~Uhlig, ``{Software-Defined Networking: A Comprehensive Survey},''
  \emph{Proceedings of the IEEE}, vol. 103, no.~1, pp. 14--76, 2015.

\bibitem{mininet-wifi}
R.~R. Fontes, S.~Afzal, S.~H.~B. Brito, M.~A.~S. Santos, and C.~E. Rothenberg,
  ``{Mininet-WiFi: Emulating software-defined wireless networks},'' in
  \emph{2015 11th International Conference on Network and Service Management
  (CNSM)}, 2015, pp. 384--389.

\bibitem{mininet}
\BIBentryALTinterwordspacing
B.~Lantz, B.~Heller, and N.~McKeown, ``{A Network in a Laptop: Rapid
  Prototyping for Software-Defined Networks},'' in \emph{Proceedings of the 9th
  ACM SIGCOMM Workshop on Hot Topics in Networks}, ser. Hotnets-IX.\hskip 1em
  plus 0.5em minus 0.4em\relax New York, NY, USA: Association for Computing
  Machinery, 2010. [Online]. Available:
  \url{https://doi.org/10.1145/1868447.1868466}
\BIBentrySTDinterwordspacing

\bibitem{Galluccio15}
L.~Galluccio, S.~Milardo, G.~Morabito, and S.~Palazzo, ``{SDN-WISE: Design,
  prototyping and experimentation of a stateful SDN solution for WIreless
  SEnsor networks},'' in \emph{2015 IEEE Conference on Computer Communications
  (INFOCOM)}, 2015, pp. 513--521.

\bibitem{inet363}
\BIBentryALTinterwordspacing
``{INET-3.6.3 Framework}.'' [Online]. Available:
  \url{https://inet.omnetpp.org/2017-12-22-INET-3.6.3-released.html}
\BIBentrySTDinterwordspacing

\bibitem{simulation}
\BIBentryALTinterwordspacing
``{Implementation of RPL in OMNeT++}.'' [Online]. Available:
  \url{https://github.com/NETSERV-UAH/RPL/tree/master/OMNeTpp5\_2\_1\_INET3\_6\_3}
\BIBentrySTDinterwordspacing

\bibitem{Rojas21}
\BIBentryALTinterwordspacing
E.~Rojas, H.~Hosseini, C.~Gomez, D.~Carrascal, and J.~{Rodrigues Cotrim},
  ``{Outperforming RPL with scalable routing based on meaningful MAC
  addressing},'' \emph{Ad Hoc Networks}, vol. 114, p. 102433, 2021. [Online].
  Available:
  \url{https://www.sciencedirect.com/science/article/pii/S1570870521000147}
\BIBentrySTDinterwordspacing

\end{thebibliography}

\end{document}